\documentclass[journal=nalefd,manuscript=letter]{achemso}

\usepackage[version=3]{mhchem} 
\usepackage{color}

\author{Yining Xuan$^\dagger$}
\author{Daito Miyazaki$^\dagger$}
\author{Yuki Ishikawa}
\author{Mark Sadgrove}
\email{mark.sadgrove@rs.tus.ac.jp}
\affiliation[TUS]
{Department of Physics, Tokyo University of Science, 1-3 Kagurazaka Shinjuku-ku Tokyo 162-8601 }

\title[]
{Chiral light from an emitter coupled to an achiral particle via the Purcell effect} 

\begin{document}


\begin{abstract} 
We demonstrate that non-chiral nanoparticles can produce chiral light when point emitters are coupled to their surface plasmon modes (SPMs) under certain conditions. Chiral emission arises from asymmetrical plasmon mode propagation from the source combined with the spin-momentum locked nature of the SPMs. The Purcell regime of cavity quantum electrodynamics (QED) ensures that radiation from the coupled mode dominates over that from the emitter itself, giving rise to photons with a circularly polarized component -- i.e. chiral light. We experimentally demonstrate this effect using electron beam-induced cathode luminescence from a gold nanorod, coupling it evanescently to a nanofiber probe which also supports spin-momentum locked light. This converts the net spin of the emission into a net directionality of propagation in the fiber modes.
\end{abstract}

The production of photons with a desired spectrum and enhanced generation rate is ably handled by cavity QED techniques, in particular the Purcell effect~\cite{Purcell, aharonovich2016solid}. By coupling photons from an emitter to a resonant mode, for which both the coupling \emph{and} mode decay occur much faster than the rate at which photons are emitted into free space, the overall emission of the combined system takes on the characteristics of the mode itself, which may be engineered by nanofabrication. 
In recent years, it has been shown that photon polarization can also be engineered if the characteristics of the resonator mode can be suitably tailored. In particular, appropriate breaking of resonator symmetries, can produce polarized emission by either large resonator birefringence~\cite{unitt2005polarization,munsch2012linearly,zhu2014polarization,pfeiffer2018coupling,chandra2020polarization,zhang2019polarized,sugawara2022plasmon,shafi2023bright}, asymmetrical pumping~\cite{chen2025observation}, or the use of chiral nanostructures~\cite{ahn2024highly,xie2025unidirectional}. Because control of polarization allows control of propagation direction in chiral quantum optics, polarization control can also be applied to the routing of photons~\cite{lodahl2017chiral,petersen2014chiral,le2015nanophotonic,xie2025unidirectional}.

It has been shown that even for an achiral particle, such as a gold nanorod (GNR), optical chirality exists in the near field~\cite{schaferling2012formation,hashiyada2018imaging,hashiyada2019active}. Nonetheless, these intriguing observations were made for effective plane-wave excitation of GNRs, and did not produce circular polarization in the far field, ostensibly due to the \emph{overall} symmetry in the near-field polarization structure. The convenience that would be afforded by being able to use a simple, non-chiral structure to induce circularly polarized photon emission from a linearly polarized source is a strong motivation for further studies in this area.

\begin{figure}
\centering
\includegraphics[width=\linewidth]{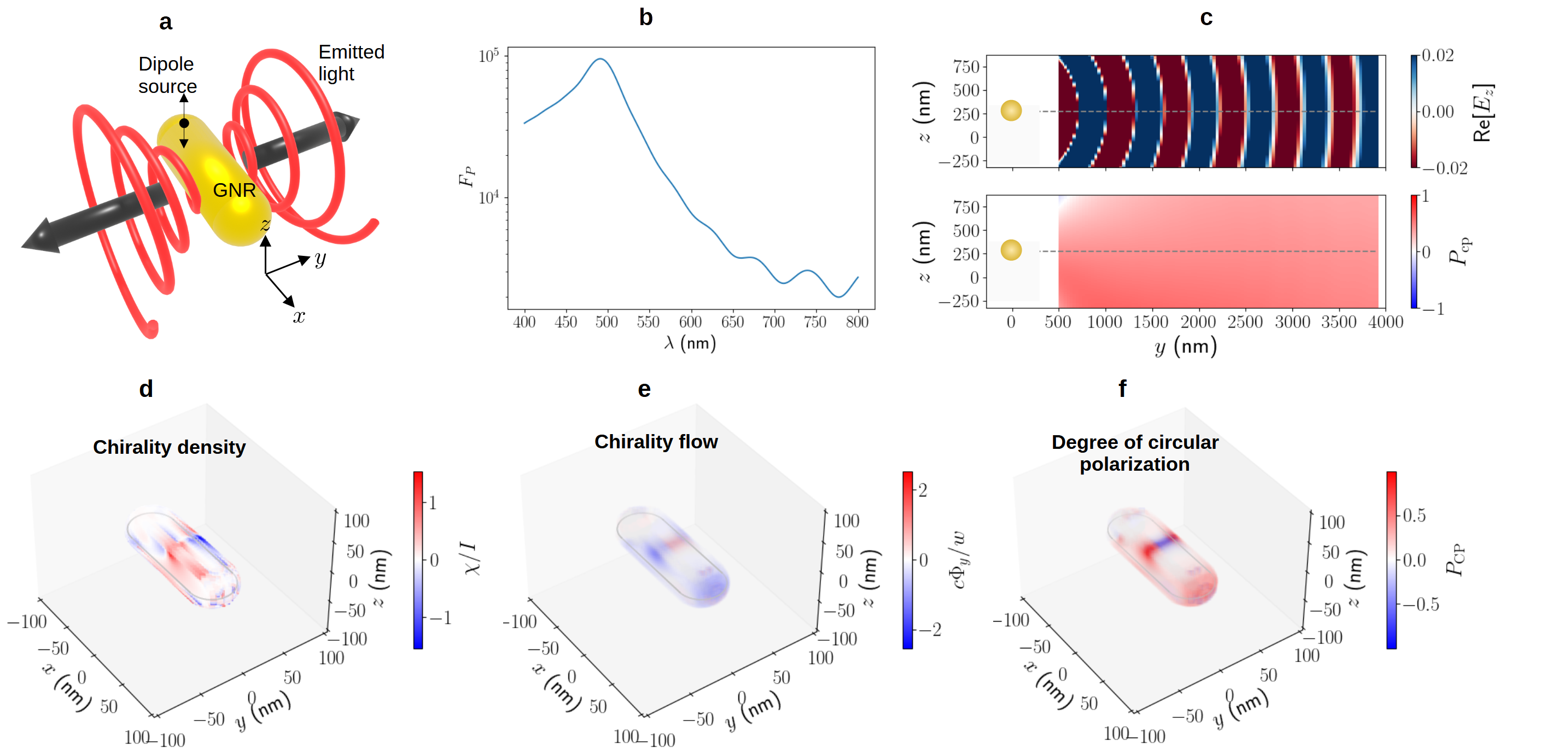}
\caption{\label{fig:1} (a) Schematic illustration of the generation of chiral light from a point emitter coupled to a GNR. (b) Purcell enhancement as a function of wavelength $\lambda$ for a dipole emitter coupled to the GNR. (c) (Upper panel) real part of the $z$-component of the electric field emitted from the GNR. (Lower panel) degree of circular polarization $P_{\rm CP}$ of the light shown in the upper panel. (d) Chirality density $\chi$ normalized by intensity. (e) $y$- component of chirality flow $\Phi_y$ normalized by energy density. (f) Degree of circular polarization $P_{\rm CP}$. }
\end{figure}
In this work, we investigate the emission produced when a point dipole is coupled to the surface plasmon modes (SPMs) of a GNR, with asymmetric emitter placement relative to the principle axes of the GNR, as depicted in Fig.~\ref{fig:1}(a).
We show that for excitation both inside and outside the GNR, this leads to an asymmetry between \emph{propagating} modes which is preserved in the SPM field distribution due to the swift decay of the mode. In addition, the large Purcell enhancement, which is well known for gold nanowire antennae and nanorods~\cite{chang2007strong,munsch2012linearly}, means that the actual emission for the coupled system comes largely from the  SPM of the GNR (Fig.~\ref{fig:1}(b)).
Then, as a consequence of spin-momentum locking of the GNR modes, the asymmetry in propagation leads to a \emph{net spin} for the dipole moment associated with the GNR, and emission  with a non-zero degree of circular polarization (DCP), as shown in Fig.~\ref{fig:1}(c). We experimentally confirm this effect by using an electron beam to induce a dipole-like excitation in a GNR, and collect the optical emission by evanescent coupling to the spin-momentum locked modes of an optical nanofiber, transforming the net DCP into a net directionality of coupled light in the fiber.

The modern understanding of optical chirality is based on the chirality density $\chi$ and the chirality flow $\mathbf{\Phi}$. For the chirality density, it is sufficient to use the definition
\begin{eqnarray}
\chi = {\rm Im}\{\mathbf{E}^*\cdot\mathbf{H}\},
\label{eq:chi}
\end{eqnarray}
for electric (magnetic) field $\mathbf{E}$ ($\mathbf{H}$) giving a quantity with units of power per unit area. 

The definition of the chirality flow in a dissipative material such as gold is somewhat subtle. We follow Alpeggiani \emph{et al.}~\cite{alpeggiani2018electromagnetic} and write
\begin{equation}
\mathbf{\Phi} = {\rm Im}\{\sqrt{\frac{\epsilon_r\epsilon_0}{\mu_0}} \mathbf{E}^*\times\mathbf{E} + \sqrt{\frac{\mu_0}{\epsilon_r\epsilon_0}} \mathbf{H}^*\times\mathbf{H}\},
\label{eq:Phi}
\end{equation}
where $\epsilon_r$ is the relative permittivity of the metal, which has a negative real part, and, in general, a non-zero imaginary part leading to dissipation. For simplicity, we ignore the imaginary part of $\epsilon_r$ when plotting $\mathbf{\Phi}$.
Note that from here on, we will consider the dimensionless chirality density $\chi / I$, where $I = 2\epsilon_0 c \mathbf{E}^*\cdot\mathbf{E}$, and the dimensionless chirality flow $\Phi /(c w)$, where $w=\epsilon_0 \mathbf{E}^*\cdot\mathbf{E} + \mu_0\mathbf{H}^*\cdot\mathbf{H}$, which are 
easier to interpret in the case of dipole excitation~\cite{yang2023inverse}. 

Lastly, we introduce a more intuitive quantity, the degree of circular polarization (DCP) $P_{\rm CP}$ which is given
by
\begin{equation}
P_{\rm CP} = \frac{|\mathbf{E}^*\cdot \mathbf{u}_L|^2 - |\mathbf{E}^*\cdot \mathbf{u}_R|^2}{|\mathbf{E}^*\cdot \mathbf{u}_L|^2 + |\mathbf{E}^*\cdot \mathbf{u}_R|^2},
\label{eq:pcp}
\end{equation}
where $\mathbf{u}_L=(\mathbf{e}_x+i\mathbf{e}_z) / \sqrt{2}$ is the polarization vector for left-hand circular polarized (LCP) light which rotates counter-clockwise in time using the phase convention $\exp[i(\mathbf{k}\cdot\mathbf{r}-\omega t)]$. The right-hand circular polarization (RCP) vector is $\mathbf{u}_R=\mathbf{u}_L^*$.

The upper panel of Fig.~\ref{fig:1}c shows a propagating electromagnetic (EM) field emitted from the GNR in the $y-z$ plane. The field is sampled at a position sufficiently far from the GNR that non-radiative field components have died down. As we show in the Appendix, the field intensity has the expected $1/r^2$ dependence of EM radiation. The lower panel of Fig.~\ref{fig:1}c shows the value of $P_{\rm CP}$ for this field. The emitted field clearly has a non-zero degree of circular polarization and thus has net optical chirality.
The radiative component of an electric field from a dipole emitter in a direction perpendicular to the dipole axis has the form~\cite{griffiths2023introduction}
\begin{equation}
\mathbf{E} = \mathbf{p}\frac{\mu_0\omega^2}{4\pi r}\exp(i(\mathbf{k}\cdot\mathbf{r} - \omega t)),
\label{eq:EM}
\end{equation}
that is, the polarization of $\mathbf{E}$ is determined by the dipole moment $\mathbf{p}$. We thus restrict our following analysis to the field inside the GNR which determines the induced dipole moment $\mathbf{p}$ and thus the chirality of the emitted radiation.

Let us now numerically establish the existence of a rotating component for the induced dipole moment in the GNR. In the point dipole approximation, let the induced polarization of the GNR be $\mathbf{p} = \alpha\mathbf{E}$. 
For the dipole moment of the GNR as an extended object, we decompose the GNR polarization as follows
\begin{equation}
\mathbf{p} = \frac{1}{V}\int_V dV \mathbf{P}(\mathbf{r}) = \frac{\alpha}{V} \int_V dV \mathbf{E}(\mathbf{r}), 
\label{eq:pol}
\end{equation}
where $V$ is the GNR volume, and $\mathbf{P}(\mathbf{r})=\alpha\mathbf{E}(\mathbf{r}$) is the polarization at point $\mathbf{r}$ within the GNR induced by the local field $\mathbf{E}(\mathbf{r})$ at that point. We assume a constant, scalar electric polarizability $\alpha$ for simplicity. Because the wavelength we consider is roughly in the middle of the plasmon resonance wavelengths for the $x$ and $z$ axes, this approximation is not unreasonable. (Note that no such approximations are made in the actual numerics.)

The local field $\mathbf{E}(\mathbf{r})$ induced by the dipole source in the GNR was found numerically using the finite-difference time-domain (FDTD) technique. We choose to place the dipole source inside the rod since this is the regime accessible experimentally, although we emphasize that the same effect exists for dipoles situated outside the rod. We then analyzed the local field chirality which, by Eq.~\ref{eq:pol}, corresponds to the chirality of the polarization. Details of the FDTD simulations regarding convergence, etc are found in the Appendix.

In Fig.~\ref{fig:1}(d), the normalized chirality density is shown over the volume of the GNR. Although the density is generally non-zero, it is crucial to note that its structure has sign flipped mirror symmetry in the $x$-$z$ plane. That is, every positive region is exactly matched by a negative region of the same size and shape under reflection. For this reason, the average value over the GNR volume vanishes. On the other hand, the distribution of $\Phi_y / (cw)$ shown in Fig.~\ref{fig:1}(e) does not have this property and has a net negative value after volume averaging. (The distributions for $\Phi_x$ and $\Phi_z$ \emph{do} have the same sign-flipped mirror symmetry as the chirality density and thus also average to zero, hence their omission here). 

Lastly, we plot $P_{\rm CP}$ in Fig.~\ref{fig:1}(f). We see it has a similar distribution to  $\Phi_y / (cw)$, with its average positive value indicating a dominant left-hand circular polarization. This also allows us to interpret $\Phi_y$ as a spin, which is negative as expected for a counter-clockwise polarization in the $x$-$z$ plane. Taken together, the results shown in Fig.~\ref{fig:1} demonstrate numerically the existence of chiral EM emission despite the fact that the dipole excitation is linearly polarized. 

\begin{figure*}
\centering
\includegraphics[width=\linewidth]{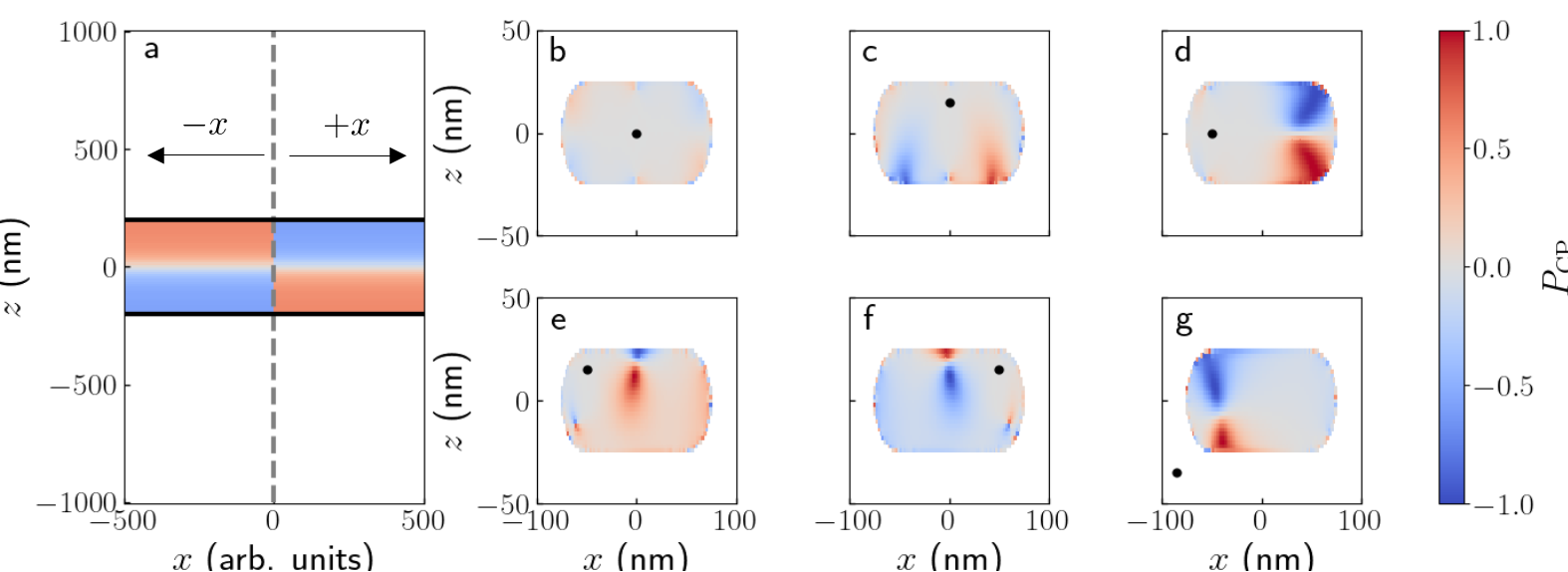}
\caption{\label{fig:2} (a) Degree of circular polarization for a gold nanowire fundamental mode at 600 nm wavelength, in the absence of dissipation. Thick black lines show the surface of the wire. The right hand half of the figure is for a $+x$ propagating mode, and the left hand half is for a $-x$ propagating mode, as indicated by the arrows. (b-f) Numerically calculated $P_{\rm CP}$ of the electric field in a GNR excited by an $z$-polarized point dipole source placed at (b) $(x=0,z=0)$,  (c) $(x=0,z=15\,\mathrm{nm})$, (d) $(x=-50 \,\mathrm{nm},z=0)$, (e) $(x=-50\,\mathrm{nm},z=15\,\mathrm{nm})$, (f) $(x=50\,\mathrm{nm},z=15\,\mathrm{nm})$, and (g) $(x=-85\,\mathrm{nm},z=-35\,\mathrm{nm})$, as shown by a black dot in each case.}
\end{figure*}

Next, we offer an interpretation of the results in Fig.~\ref{fig:1} in terms of the polarization properties of the modes of a metal nanowire. Having verified the three dimensional distribution of the three chirality measures in Figs.~\ref{fig:1}(d-f), we now restrict our attention to the $x$-$z$ plane for ease of visualization. First, we review the properties of the lowest order mode of a nanowire - which has a TM$_{01}$ character~\cite{takahara1997guiding, chang2007strong}.
The $x$ and $z$ components of the field in the $x$-$z$ plane are 
\begin{eqnarray}
E_x &=& E_0 I_0(\gamma z),\\
E_z &=& -i\frac{\beta}{\gamma}E_0 I_1(\gamma z),
\end{eqnarray}
where $I_n$ are modified Bessel functions of the first kind of order $n$, $\gamma$ is the mode transverse wave number, and $\beta$ is the mode propagation constant, which is positive (negative) for positive (negative) $x$ propagation.

From this definition, we see from comparison to the circular polarization vectors that a $+x$ propagating mode has left-hand circular polarization component for $z>0$, and right-hand circular polarization component for $z<0$ due to the fact that $I_1$ is an odd function. The situation is reversed for negative $\beta$, as shown in Fig.~\ref{fig:2}(a).
Specifically, for the wavelength of 600 nm used here, and a wire diameter equal to that of the GNR (50 nm), we have
$\beta / \gamma = 0.58$, and a transverse-to-longitudinal field ratio $|E_z| / |E_x|$ at the wire surface of 0.29. (This ratio is 1 for perfect circular polarization). 

Moving on to simulations of the field created by a dipole source, Fig.~\ref{fig:2}(b) shows the degree of circular polarization associated with the plasmon mode created by a dipole in the center of the GNR. We note that it is qualitatively the same as that seen for a nanowire, implying that the distribution of polarization within the GNR is due to propagating modes with spin-momentum locking. An objection to this point of view is that a GNR supports \emph{localized} surface plasmon modes i.e. modes with a standing wave character, and thus the spin-momentum locking concept does not apply. However, the extremely fast decay time of the SPM, which is only slightly more than its round trip time inside the GNR (see Appendix), means that the linearly polarized localized mode is not established within the lifetime of the excitation. Thus the polarization state of the propagating plasmon excitation dominates, and determines the polarization of the local induced dipole moment as seen in Fig.~\ref{fig:2}(b). Figures~\ref{fig:2}(c) and (d) show what happens to the distribution of polarization when the dipole is displaced along the $x$ or $z$ axis respectively. Unsurprisingly, this breaks the symmetry due to the longer traversal length in one direction.

Now, the examples shown in Figs.~\ref{fig:2}(b-d) all show sign flipped mirror symmetry with respect to at least one plane and thus there is zero degree of circular polarization when averaged over the GNR volume. However, if the dipole is displaced from the center along \emph{both} the $x$ and $z$ axes, then a single lobe with well-defined $P_{\rm CP}$ will dominate, as seen in Figs.~\ref{fig:2}(e) and (f). Here, the broken symmetry of mode propagation leads to the existence of a net DCP of the electric field inside the GNR, and, thus, in the induced dipole moment. Therefore, by Eq.~\ref{eq:EM}, the EM radiation emitted from the GNR also has a degree of circular polarization. In addition, Fig.~\ref{fig:2}g demonstrates that the effect also holds for a dipole coupled from outside the GNR. 

The effect noted here is transient in the sense that if the oscillation continued for more time, or if the GNR was longer and the mode propagated over a longer distance, the mode would eventually settle into an eigenmode with no net spin. However, due to the extremely short SPM lifetime and the short length of the GNR relative to the wavelength, this ``transient" field distribution dominates, leading to a net polarization of the SPM over its lifetime. Further tests of this concept are found in the Appendix. 
\begin{figure}
\centering
\includegraphics[width=\linewidth]{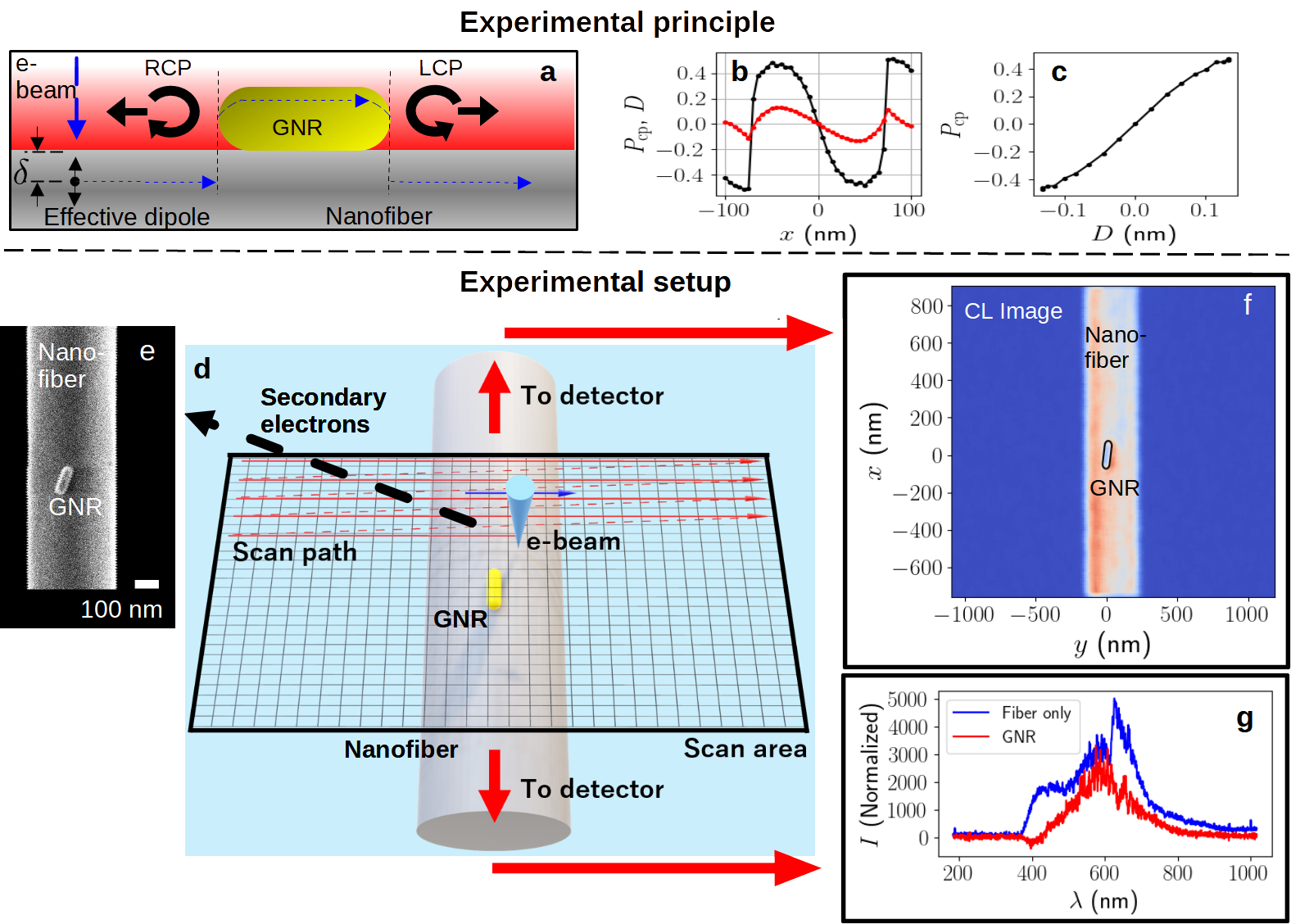}
\caption{\label{fig:3} (a) Principle of the experiment. An electron beam is incident on the GNR, creating an effective dipole excitation at the position where the electron stops. CL from the so-excited GNR plasmon mode couples to the optical nanofiber fundamental mode with a directionality dependent on the circular polarization component of the induced dipole moment.
(b) Simulated values of $P_{\rm CP}$ (black circles) and $D$ (red circles) as a function of emitter position $x$. (c) Dependence of $P_{\rm CP}$ on $D$ inside the GNR, taken from the data shown in (b). 
(d) Experimental setup inside the SEM. A GNR deposited on an optical nanofiber is excited by an electron beam (diameter $\sim 5$ nm). The resulting optical emission is coupled evanescently to the fiber, and sent via a fiber feedthrough to a detector. (e) SEM image of the GNR and nanofiber from secondary electron detection. (f) CL image from light detected at one end of the fiber using a single photon counting module (SPCM) (g) Spectrum of the bare fiber (blue line) and the GNR (red line), as indicated in the legend, measured by an optical multi-channel analyzer.}
\end{figure}

We now move on to our experimental investigation of the effect described above. 
We use the cathode luminescence (CL) method to excite emission from the GNR using a beam of electrons inside a scanning electron microscope (SEM) which function as point dipole sources when entering or passing near the GNR. It is well known that the CL signal is proportional to the local photonic density of states, which is in turn proportional to the Purcell factor for an emitter coupled to the GNR plasmon
~\cite{garcia2010optical,polman2019electron, garcia2021optical}.
As shown in Fig.~\ref{fig:3}(a), rather than standard collection of emitted light in the far field, we couple it to the evanescent field of an optical nanofiber (ONF) aligned with the GNR axis. Due to the well known spin-momentum locking of the ONF's evanescent field~\cite{petersen2014chiral,lodahl2017chiral}, this results in the optical chirality being converted to a difference in coupled intensity to the positive and negative $x$ propagating modes of the fiber. 

We define the directionality of the light coupled from the GNR dipole moment to the fiber as 
$D=\frac{I_1-I_2}{I_1 + I_2}$,
where $I_1$ is the $+x$ propagating intensity and $I_2$ is the $-x$ propagating intensity.
Now, the intensity of coupling to a given mode $\epsilon$ is given by~\cite{novotny2012principles} $I\propto|\mathbf{p}\cdot\mathbf{\epsilon}|^2$. Using $\epsilon(\mathbf{r})_\pm$ to represent the $\pm x$ propagating fiber modes, we find, at a given point $\mathbf{r}$,
\begin{equation}
D(\mathbf{r}) = \frac{|\mathbf{P}\cdot\epsilon_+|^2 - |\mathbf{P}\cdot\epsilon_-|^2}{|\mathbf{P}\cdot\epsilon_+|^2 + |\mathbf{P}\cdot\epsilon_-|^2} = \frac{|\mathbf{E}\cdot\epsilon_+|^2 -|\mathbf{E}\cdot\epsilon_-|^2}{|\mathbf{E}\cdot\epsilon_+|^2 + |\mathbf{E}\cdot\epsilon_-|^2},
\label{eq:D}
\end{equation}
where we applied Eq.~\ref{eq:pol} to reach the final form. Equation~\ref{eq:D} is equivalent to Eq.~\ref{eq:pcp} for the degree of circular polarization if the spin-momentum locked fiber modes are perfectly circularly polarized. In reality, the modes have elliptical polarizations with a weaker $x$ component~\cite{le2004field}. Nonetheless, 
the two measures have the same structure as seen in Fig.~\ref{fig:3}(b), in particular peaking at the same $x$ value. 
In Fig.~\ref{fig:3}(c) 
we plot $P_{\rm CP}$ as a function of $D$ within the GNR. The monotonicity of the relationship shows that the directionality functions as a proxy for the circular polarization of the emitted light.

Having established the principle of our technique for measuring $P_{\rm CP}$ we move on to the experimental setup. Figure~\ref{fig:3}(d) shows a schematic representation of our experiment. A tapered optical fiber with a waist diameter of $\sim 500$ nm is mounted in the sample chamber of a scanning electron microscope (SEM, Carl  Zeiss SUPRA50). 

A single GNR (Nanopartz A12-50-808, diamete 50 nm, length 150 nm) is deposited on the nanofiber surface with a micro-pipette~\cite{sugawara2020optical}. 
A fiber feedthrough allows the collected CL to be measured using single photon counting modules (SPCMs) located outside the SEM~\cite{uemura2021probing}, as shown in Fig.~\ref{fig:3}(f). Aside from CL measurement by SPCM, it is also possible to take simultaneous measurements of secondary electrons to produce a standard SEM image (Fig.~\ref{fig:3}(e)) and measurements of the CL spectrum using an optical multi-channel analyzer (Kymera 193i, Newton DU970P-BVF). Fig.~\ref{fig:3}(g) shows recorded spectra for the fiber alone (black curve) and for the GNR with the fiber background subtracted (red curve). We see that the peak CL wavelength occurs near 600 nm, and thus this wavelength was used in our simulation results, shown so far. 

\begin{figure}
\centering
\includegraphics[width=0.9\linewidth]{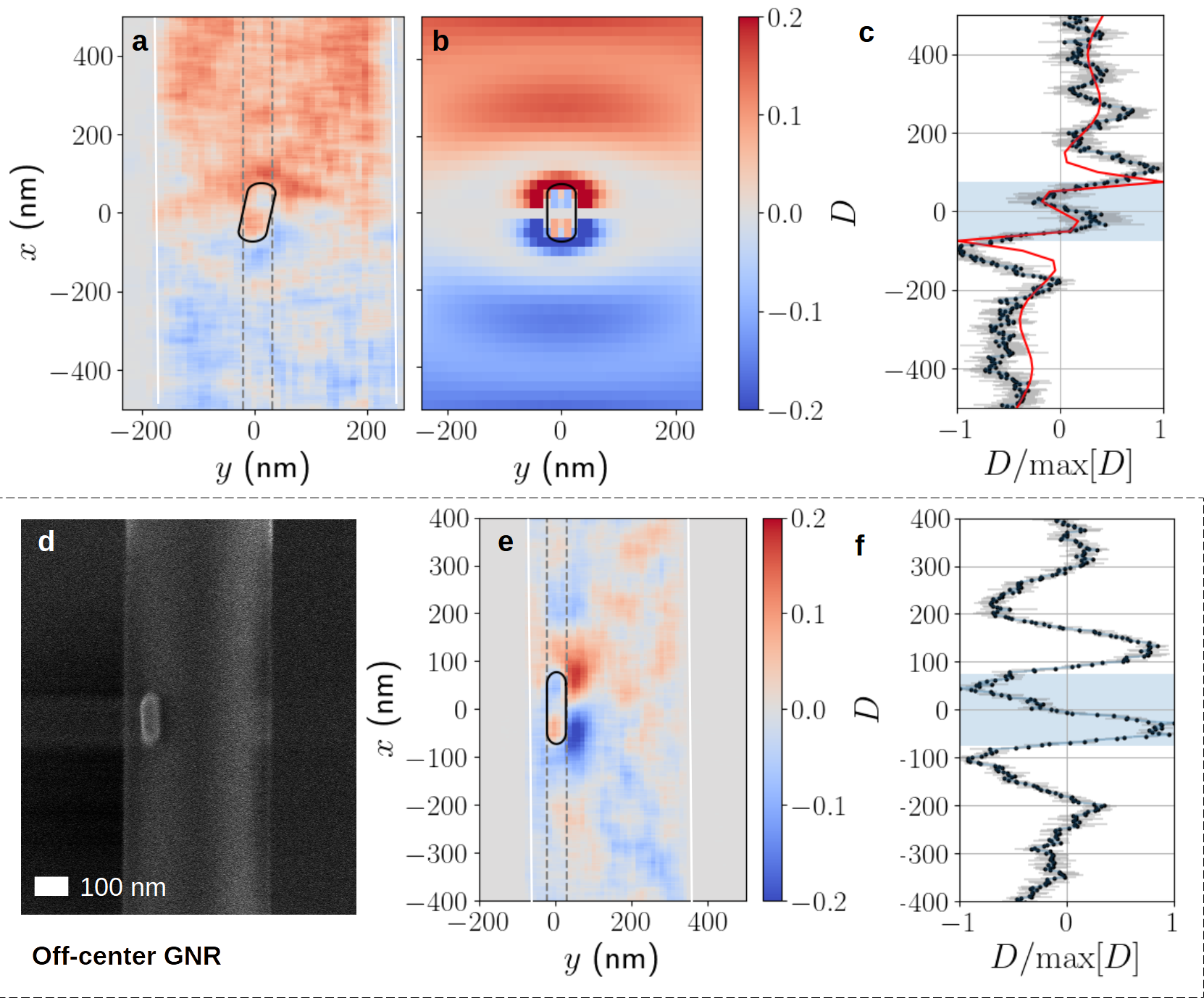}
\caption{\label{fig:4} Experimental measurements of $D$ for the GNR shown in Fig.~\ref{fig:3}e. (a) Experimentally measured directionality calculated from the data shown in Fig.~\ref{fig:3}d. Vertical white lines indicate the fiber edges. (b) Numerically calculated directionality. (c) Comparison of experimentally measured directionality integrated over the GNR width (black dots with gray error bars) with the numerically predicted value (red curve). (d-f) Similar measurements showing the effect of displacement of the GNR from the ONF center. (d) SEM image of the GNR sample used. (e) The measured directionality for the sample shown in (d). (f) The averaged directionality over the region indicated by the gray dashed lines in (e). In both cases the pale blue regions in (c) and (f) indicate the GNR region. }
\end{figure}
We now turn to the main result of the paper - the demonstration of emission with a circular polarization component from a GNR using spin-momentum locked CL measurements. Figure~\ref{fig:4}(a) shows the directionality calculated from the CL measurements shown in Fig.~\ref{fig:3}(d) in a region surrounding the GNR. The associated simulation results are shown in Fig.~\ref{fig:4}(b). 

The experimental result demonstrates a number of features predicted by the detailed numerical simulations. First, there is the directionality inside the GNR which, as predicted, flips from positive to negative as $x$ is increased past the origin, owing to a change in the dominant circular polarization from LCP to RCP. 

Second, the structure of $D$ outside the GNR is also reproduced. Most intriguingly, the reversal of directionality when the GNR boundary is crossed along with oscillations in $D$ may be seen. These details are more clearly seen by looking at the behavior of $D$ as a function of $x$ with $y = 0$, as shown in Fig.~\ref{fig:4}(c). To make this graph, the experimental data in Fig.~\ref{fig:4}(a) was averaged over the region indicated by the gray dashed lines, and normalized to its maximum giving the black points shown, while the light gray region shows the standard deviation over the same region. The red curve in Fig.~\ref{fig:4}(c) shows the normalized simulation results corresponding to $y=0$. We see good correspondence between the predicted and measured structure in $D$, including the reversal of sign, and position of peaks. 

The directionality seen outside the GNR in both simulations and experiments must be interpreted with some care compared to that seen inside the GNR. This is because the source of the dipole excitation outside the GNR is the excitation of non-bridging oxygen hole centers (NBOHCs) in the fiber~\cite{uemura2021probing} which couple to the GNR plasmon, but also experience interference due to scattering from the GNR. This leads to the oscillations in $D$ seen outside the GNR. We also note that the center wavelength of NBOHCs in silica is about 650 nm, in comparison to the 600 nm center wavelength seen for the GNR CL spectrum. This wavelength difference is taken into account in our simulations.

A natural question is how the relative position of the GNR and fiber affects the results. In  Figs.~\ref{fig:4}(e),(f), we show measurements of $D$ for a GNR which is near to the edge of the ONF, as seen in the SEM image in Fig.~\ref{fig:4}(d). Although the distribution of $D$ around the GNR is distorted, the same qualitative behavior seen in Figs.~\ref{fig:4}(a-c) is reproduced.
We note that our experimentally measured values of $D$ approach the numerically calculated limit of 0.15 which corresponds to a value of $P_{\rm CP}$ of the emitted light of about 0.4. 
Although a systematic study is beyond the scope of the present work, we have found numerically that for a longer wavelength, the value of $P_{\rm CP}$ can rise to almost unity. We give an example in the Appendix.

In conclusion, we have numerically investigated the phenomenon of chiral light produced from non-chiral GNRs when a dipole source is coupled to the GNR at a position displaced from its center. The emergence of a net chirality can be understood in terms of the excitation of spin-momentum locked modes in the GNR. We tested this phenomenon experimentally by using an electron beam to produce an effective point dipole excitation in a GNR and collected the induced cathode luminescence evanescently. This allowed the polarization to be converted to directionality of mode propagation, due to the spin-momentum locking of the nanofiber modes.

Although our results might seem at odds with established facts regarding the linear polarization of emitters coupled to anisotropic resonators~\cite{unitt2005polarization,munsch2012linearly,zhu2014polarization,pfeiffer2018coupling,chandra2020polarization,zhang2019polarized}, to the best of our knowledge, no previous studies have had the necessary control over emitter-particle positioning or the right particle geometry to discover the effect reported here. In addition, although electron beam methods have been used to induce directional polarization from isotopic particles~\cite{coenen2014directional}, this was not mode-resolved or correlated with circular polarized emission, as in the present study.

Generation of circularly polarized photons by coupling an emitter to a non-chiral nanostructure greatly simplifies the generation of chiral light. Furthermore, we note that the realization of directionality of emission is also useful in itself, and has been the focus of many studies in the field of chiral quantum optics in recent years~\cite{lodahl2017chiral}. Finally, we note that evanescent collection of CL, used here for the first time, is promising, as it sheds light on new aspects of cathode luminescence through the spin-momentum locking property. In this case, it allowed us to use directionality as a proxy for light chirality, which would have been difficult to measure with standard CL setups, even those equipped with polarization analysis.


\subsection*{Author contributions}
$^\dagger$YX and DM contributed equally to this work.
YX performed numerical simulations and discovered the circularly polarized emission phenomenon. DM performed the experiment and discovered the flip in directionality inside the GNR. YI contributed to data analysis and the development of the experimental technique. MS supervised the project and contributed to simulations and the overall understanding of the observations. All authors contributed to the writing of the paper.
\subsection*{Notes}
The authors declare no competing financial interest.

\appendix
\section{APPENDIX}
\renewcommand{\thesection}{A\arabic{section}}   
\renewcommand{\thefigure}{A\arabic{figure}}
\setcounter{figure}{0}
\renewcommand{\thetable}{A\arabic{table}}
\setcounter{table}{0}

\section{Simulation methods}
In this section we give details regarding our finite difference time domain (FDTD) simulations.
All simulations were performed using a commercial FDTD solver~\cite{Lumerical}.  
\subsection{Simulation setup}
In all our simulations, the gold nanorod is a hemisphere capped cylinder of total length 150 nm and radius 25 nm.
The refractive index is set to the provided preset which uses complex index values taken from the CRC handbook of Chemistry and Physics~\cite{haynes2016crc}.

The gold nanorod is surrounded by a custom mesh area of volume 1 $\mu$m$^3$ in which a mesh size of 2 nm is set. Outside this area, the mesh size is left to the software to determine by setting a preset mesh level of 8. The source used for the simulations is a numerical approximation of a $z$-polarized point dipole, with a wavelength centered on 600 nm for excitation within the GNR. The placement is decided by the penetration depth $\delta$ of 2 keV electrons into a gold surface which we take as 10 nm~\cite{zarraoa2019imaging}. 

For simulations which include a nanofiber, the fiber material is set to a silica preset, with a refractive index of approximately 1.45 near 600 nm. The dipole position is decided by the penetration depth of 2 keV electrons into a silica surface which we take as 20 nm~\cite{uemura2021probing}. 

Simulations were typically run until the remaining electromagnetic energy dropped to $0.001\%$ of the initial energy. 

\section{Supporting simulation results - checks of simulation validity}

\subsection{Convergence of results with mesh density}

In FDTD simulations, verifying the convergence with respect to mesh density is an essential step to ensure the reliability of numerical results. We systematically refined the mesh density to examine its influence on the calculated transmission and directionality. In each case, the dipole source was placed in the same position in the upper-right corner of the GNR. This configuration excites an RCP component in the dipole moment, which preferentially couples into the $-x$ direction, as recorded by power monitors placed along the $\pm x$ directions.

\begin{figure}
\centering
\includegraphics[width=0.9\linewidth]{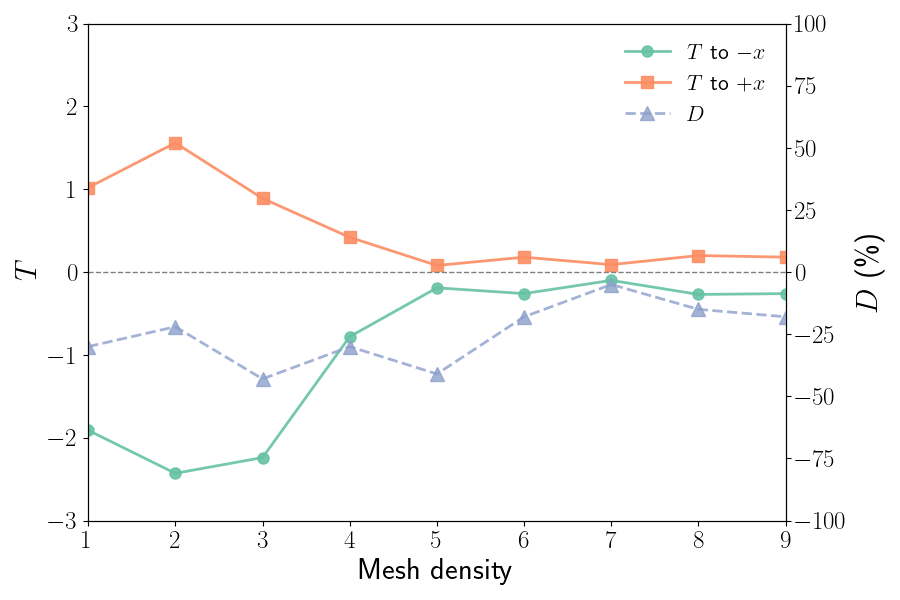}
\caption{\label{sfig:mesh_density} Transmission and directionality for a $145 \,\mathrm{nm} \times 50 \,\mathrm{nm}$ GNR plasmon mode excited by a point dipole source located at the upper-right corner under different mesh densities. Levels 1–8 correspond to the default automatic mesh settings in FDTD (from coarsest to finest). Level 9 corresponds to mesh size 8 with an additional refined custom mesh of $2\,\mathrm{nm}$ applied around the GNR.}
\end{figure}

The results are shown in Fig.~\ref{sfig:mesh_density}. Here, the horizontal axis (mesh density levels 1–8) corresponds to the default automatic mesh sizes defined in the FDTD software, where level 1 is the coarsest and level 8 the finest. Mesh density level 9 denotes a customized setting, in which the overall mesh was kept at level 8 while an additional refined mesh of $2\,\mathrm{nm}$ was applied locally in a 1 $\mu$m$^2$ volume centered on the GNR, as used in the simulations presented in the main paper. Although fluctuations in the parameters are not completely eliminated, the transmissions are seen to reduce steadily for mesh sizes between 2 and 5 after which they are distributed over a small range of values. Notably, the sign of the directionality is always negative as expected, but declines with refinement, saturating to a value close to 20$\%$ This demonstrates satisfactory convergence and confirms the reliability of our simulation results.

\subsection{Scaling of emitted electric field}
\begin{figure}
\centering
\includegraphics[width=0.9\linewidth]{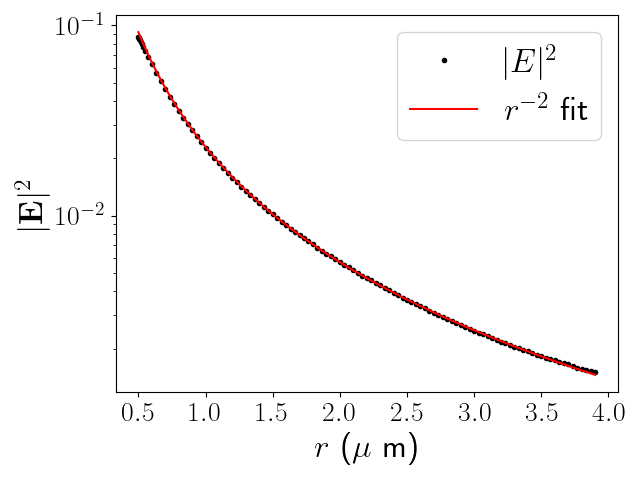}
\caption{\label{sfig:scaling} The simulated electric field intensity $|E|^2$ is plotted as a function of the distance from the source $r$ along the y axis (black dots). The red line shows a fit of the function $ar^{-2} + b$ to the numerical data.}
\end{figure}

In Fig.~\ref{sfig:scaling}, we present data showing that the numerically predicted fields correspond to radiation
by showing that the field intensity follows an inverse square law. In this case, we have plotted the field
along the $y$-axis. The results show an excellent fit of the function $ar^{-2}+b$ to the data, where $a$ and $b$ are constants, and $r$ is the distance from the GNR center along the $y$-axis. This demonstrates that the calculated field in this region corresponds to emission as expected.

\subsection{Direct observation of rotating polarization in simulations}
\begin{figure}
\centering
\includegraphics[width=0.8\linewidth]{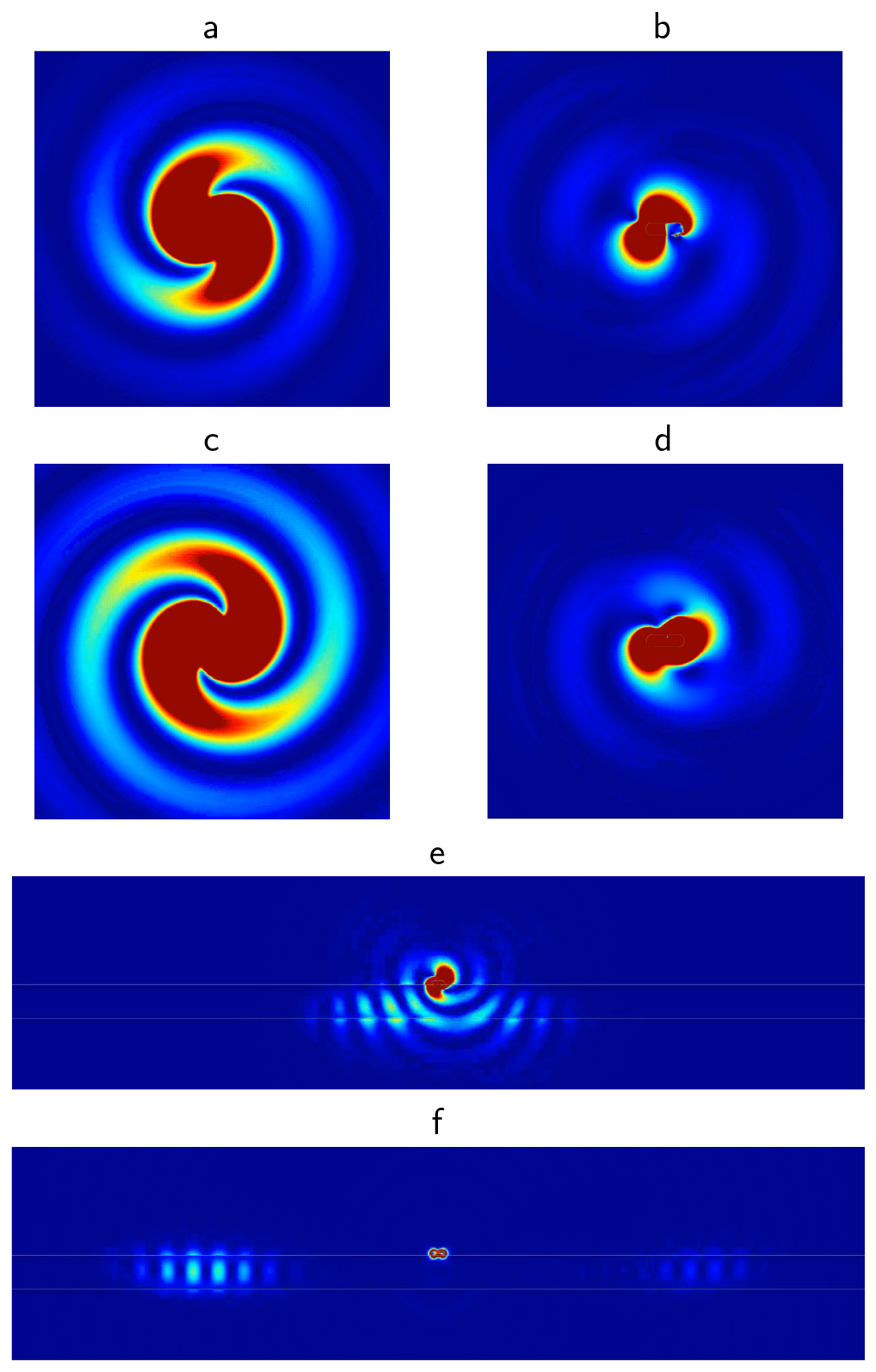}
\caption{\label{sfig:movie} Snapshots from the FDTD movie monitor. (a) and (c) correspond to the ideal LCP and RCP cases constructed using two orthogonal dipole sources with a $\pm90^\circ$ phase difference. (b) and (d) show results for a GNR with dimensions $150\,\mathrm{nm} \times 50\,\mathrm{nm}$. (e) and (f) show the case where the GNR is positioned on the surface of an ONF, and for which the dipole source was placed at the upper-right corner. (e) shows the initial generation of light with an RCP component, while (f) demonstrates the subsequent directional coupling into the ONF via spin–momentum locking.}
\end{figure}

Circular polarization may be observed directly in the simulations by recording the fields at each time step, creating a movie. Representative snapshots under different conditions are presented in Fig.~\ref{sfig:movie}.

As a reference, we first constructed ideal cases of left- and right-handed circular polarization (LCP and RCP) by simulating two orthogonal point dipole sources with a $\pm90^\circ$ phase difference. The corresponding results are shown in Figs.~\ref{sfig:movie}a (LCP) and c (RCP). When the point dipole source is placed at the \textit{upper-left} corner of the GNR, light with an LCP component is generated, as seen in Fig.~\ref{sfig:movie}b. Similarly, when the dipole source is located at the \textit{upper-right} corner, an RCP component appears, as shown in Fig.~\ref{sfig:movie}d.


Figs.~\ref{sfig:movie}(e) and (f) present the case where the GNR was placed on the surface of an ONF, with the point dipole source positioned at the \textit{upper-right} corner. According to the mechanism discussed in the main text, light with an RCP component is generated, as shown in Fig.~\ref{sfig:movie}(e). Subsequently, due to spin–momentum locking, the excitation couples directionally to the ONF as seen in Fig.~\ref{sfig:movie}(f).

\section{Supporting simulation results - further investigation of the rotating dipole moment effect}

In the following subsections, we provide further numerical evidence of the main finding of our paper - the existence of an induced
rotating polarization which gives rise to chiral emission and directional coupling to the optical nanofiber. 

Let us first qualitatively recap the understanding of this phenomenon presented in the main manuscript. We claim that a linearly polarized dipole
emitter couples to a GNR in such a way that a rotating induced dipole moment is created within the GNR. The fundamental source of the rotating
polarization is the spin momentum locked nature of the surface plasmon modes which propagate from the position of the dipole.
Now, these modes do not carry net spin when averaged over the wire volume. However, as we show in the main paper, 
if the dipole position is shifted with respect to both the principle axes, the resulting asymmetry in propagating gives rise to a dominant lobe with a certain,
non-zero degree of spin, which remains even after volume averaging.

For the average spin induced in this way to be non-trivial after averaging, the following conditions must hold.
\begin{enumerate}
\item The plasmon mode decay time must be short enough that the plasmon mode does not have time to settle into a linearly polarized eigenmode of the GNR.
\item The rod must be short enough that the initial ``transient" distribution created by the asymmetry of the source placement does not 
get outweighed by the wire eigenmode which the system will settle into after sufficient propagation length
\end{enumerate}

The following simulations test some aspect of the above conditions, after first establishing basic parameters of the GNR modes.

\subsection{GNR propagating mode properties}
\begin{figure}
\centering
\includegraphics[width=0.9\linewidth]{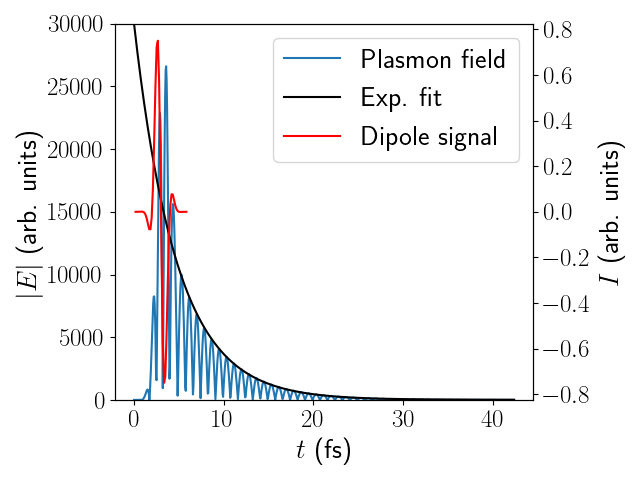}
\caption{\label{sfig:time_response} Time response of GNR plasmon mode after excitation by a point dipole source. The source time signal is shown in red.}
\end{figure}

In Table~\ref{tab:wiremode}, we summarize the properties of the fundamental propagating mode in the GNR,
along with properties related to the localized surface plasmon resonance of the GNR. These properties were determined
by excitation with an effective $\delta$ function pulse, whose time dependence is shown in Fig.~\ref{sfig:time_response}
\begin{table}[htb]
\begin{tabular}{|l|l|l|}
\hline
\textbf{Parameter}              & \textbf{Symbol or expression}     & \textbf{Value}                                                               \\ \hline
Material                        & -                   & Gold                                                                         \\ \hline
GNR length                      & $L$                   & 150 nm                                                                        \\ \hline
GNR radius                      & $a$                   & 50 nm                                                                        \\ \hline
Mode wavelength                 & $\lambda$           & 600 nm                                                                       \\ \hline
Core dielectric const. @ 600 nm & $\epsilon_{\rm co}$ & -8.4398                                                                      \\ \hline
Mode effective index                 & $n_{\rm eff}=\beta / k$       & 2.08                                                                         \\ \hline
Plasmon lifetime                   & $\tau$              & 4.75 fs                                                                      \\ \hline
Plasmon decay rate                 & $\gamma$            & 210 THz                                                                      \\ \hline
Mode round trip time       & $T_L=2n_{\rm eff}L/c$         & 2.08 fs                                                                      \\ \hline
\end{tabular}
\caption{\label{tab:wiremode} Properties of the lowest order mode of a nanowire with the same radius as the GNR. }
\end{table}


The response of the GNR plasmon mode was calculated as depicted in Fig.~\ref{sfig:time_response}

\subsection{Dependence of results on dipole duration}

The effective lifetime of the pulses used to excite the system in FDTD simulations is similar to or shorter than the plasmon lifetime of the GNR. In realistic situations, assuming the Purcell regime, the reverse is actually the case. Our interpretation of the phenomenon observed here in terms of propagating modes of the GNR does not explicitly invoke the time-domain behavior of the source, and thus, if it is correct, the results should not depend on the lifetime of the excitation. On the other hand, a priori, it seems possible that the circular polarization effect could be transient, and that for sources with lifetimes longer than that of the plasmonic mode, the circular polarization component could vanish, as the system settles into a linearly polarized eigenmode.

\begin{figure}
\centering
\includegraphics[width=0.9\linewidth]{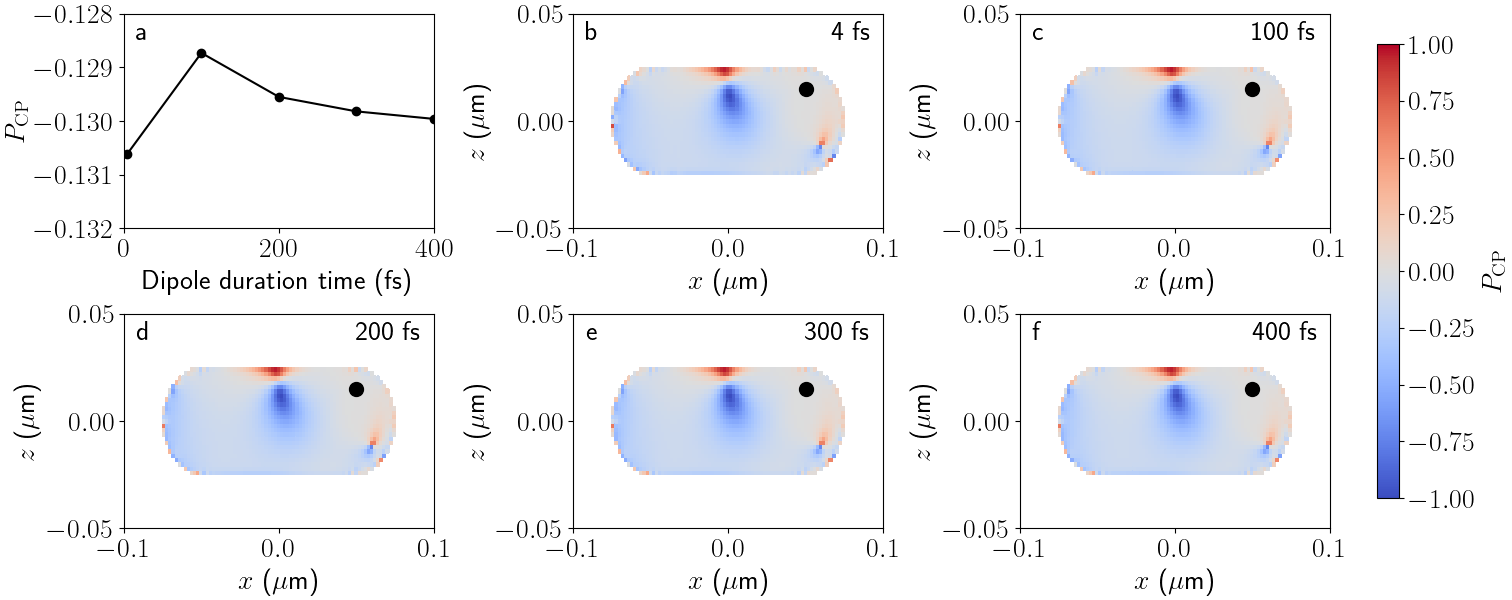}
\caption{\label{sfig:dipole_duration} (a) Degree of circular polarization $P_{\rm CP}$ and (b–f) spatial distributions inside a $150 \,\mathrm{nm} \times 50 \,\mathrm{nm}$ GNR plasmon mode excited by a point dipole source located at the upper-right corner. Results are shown for different dipole durations: (b) optimized short pulse ($\sim4 \,\mathrm{fs}$), (c) $100 \,\mathrm{fs}$, (d) $200 \,\mathrm{fs}$, (e) $300 \,\mathrm{fs}$, and (f) $400 \,\mathrm{fs}$.}
\end{figure}

To check this, we varied the duration time of the point dipole source in the FDTD simulations. The default optimized short pulse has a duration of approximately $4 \,\mathrm{fs}$, i.e. of the same order as the plasmon lifetime. In addition, we performed simulations with longer durations of $100$, $200$, $300$, and $400 \,\mathrm{fs}$, i.e. up to approximately two orders of magnitude longer than the plasmon lifetime. The results for a point dipole source positioned at the upper-right corner, are shown in Fig.~\ref{sfig:dipole_duration}.

As illustrated in Fig.~\ref{sfig:dipole_duration}a, the $P_{\rm CP}$ over the rod volume is almost unaffected by the dipole duration. Furthermore, Figs.~\ref{sfig:dipole_duration}b–f show that the spatial distribution of the DCP also remains essentially unchanged for different dipole source durations.

\subsection{Effect of optical loss on DCP of the dipole moment}

A key point of our interpretation of the emission of CP light from the emitter-GNR system is that the rapid decay of the plasmon excitation allows the rotating polarization of the spin-orbit coupled propagating nanowire mode to dominate over the linearly polarized eigenmodes of the GNR. 

To test this concept, we removed ohmic losses from the nanorod by making its permittivity purely real and negative (i.e. by making its refractive index pure imaginary). 
In this dielectric model, the complex refractive index of the nanorod is expressed as $\tilde{n} = n + i\kappa$, where $ n $ and $ \kappa $ denote the real and imaginary parts of the refractive index, respectively.  The corresponding complex permittivity is given by $ \varepsilon = \tilde{n}^2 = (n + i\kappa)^2 = (n^2 - \kappa^2) + i(2n\kappa)$. As \( n \) increases, the real part of $ \varepsilon $ decreases, leading to the introduction of an imaginary component of the permittivity and thus optical loss~\cite{takahara1997guiding}. Under our interpretation, increasing loss should lead to increasing domination of the circularly polarized component of the dipole moment as the system does not have time to settle into the linearly polarized eigen modes of the system.

\begin{figure}
\centering
\includegraphics[width=0.9\linewidth]{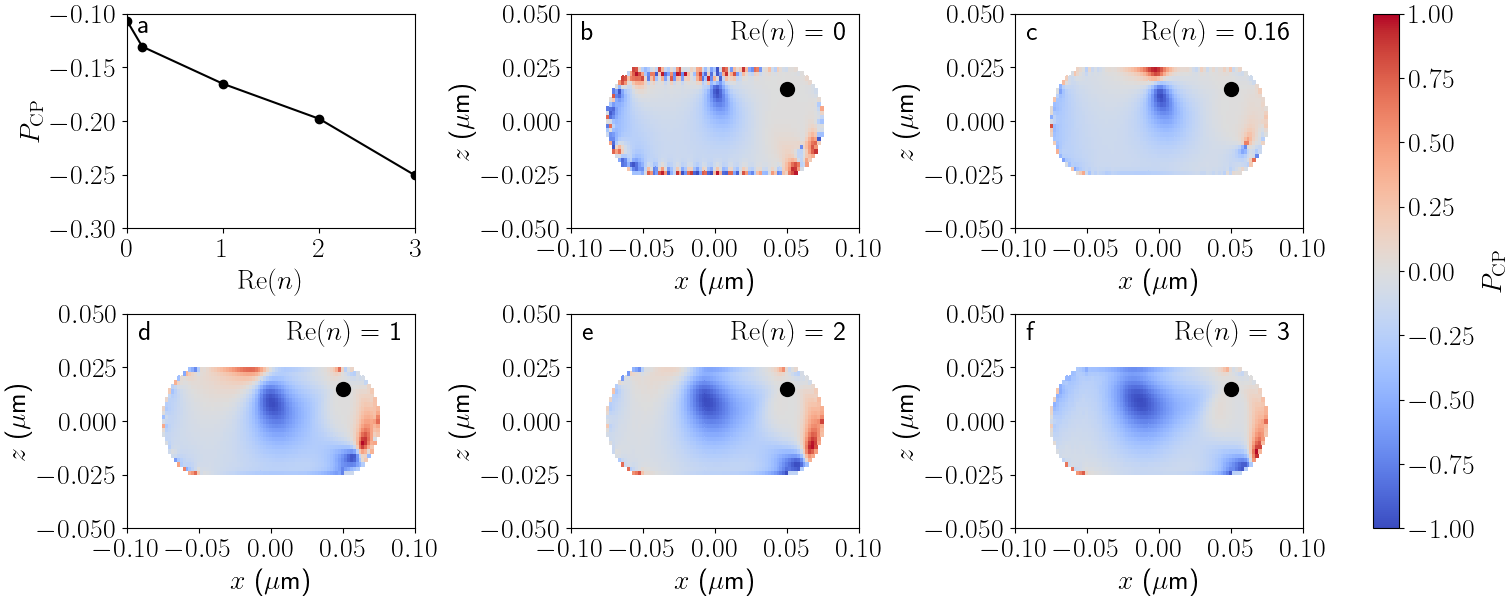}
\caption{\label{sfig:refractive_index} (a) Degree of circular polarization $P_{\rm CP}$ for different real parts of the refractive index. (b–f) spatial distributions inside the GNR when the real part of the refractive index of gold at 600 nm is set to $0$, $0.16$ (CRC reference), $1$, $2$, and $3$, respectively. }
\end{figure}

In the simulation, a dipole source was placed near the upper-right corner inside the GNR.  We systematically varied the real part of the refractive index of the GNR material, while keeping the imaginary part and other optical parameters fixed to the CRC gold values at $\lambda = 600~\mathrm{nm}$.  
The real part of the refractive index was set to $ n = 0, 1, 2, 3 $, as well as the CRC reference value as illustrated in Fig.~\ref{sfig:refractive_index}, to examine the effect of the end reflection on the CP emission. As can be seen, the results are in qualitative agreement with expectations under our interpretation. However, it is important to note that without optical loss, the mode can in principle oscillate indefinitely, whereas our simulations are necessarily for a limited time.
Thus, the results for $n=0$ are an overestimate of the DCP which in principle should tend to zero as the simulation time increases.


\subsection{Effect of GNR aspect ratio on DCP of GNR dipole moment}

\begin{figure}[htbp]
\centering
\includegraphics[width=0.9\linewidth]{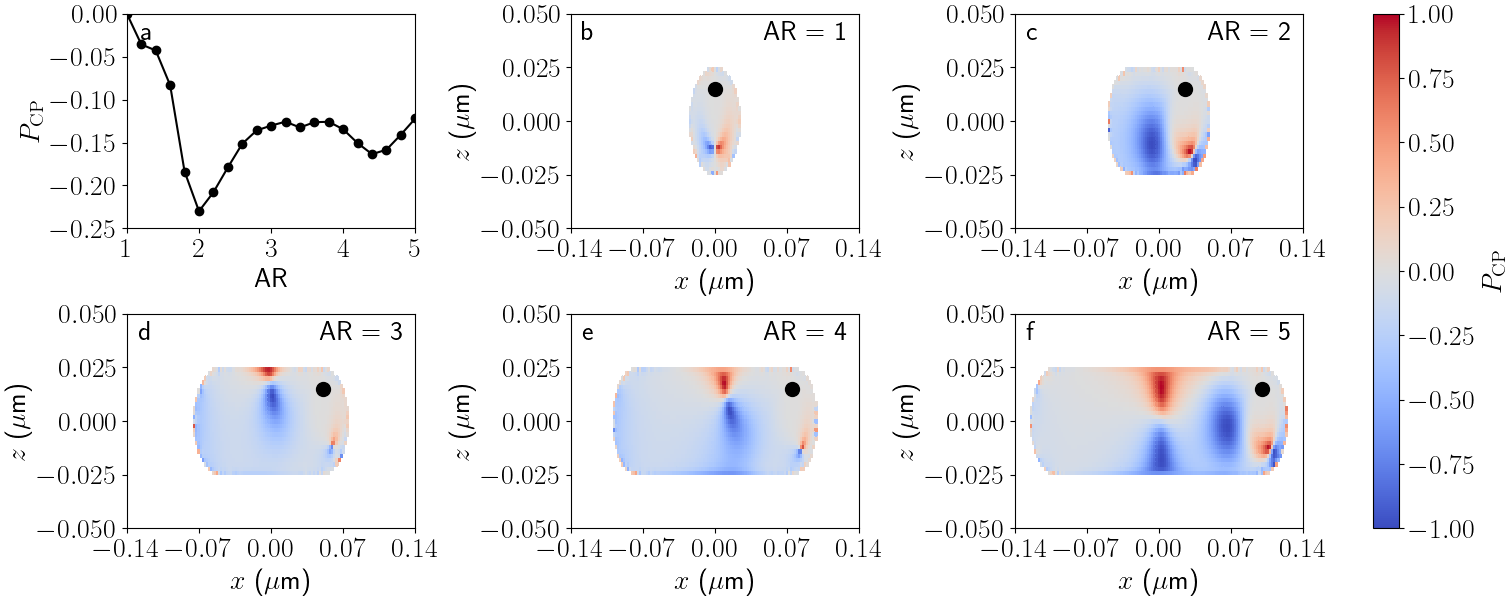}
\caption{\label{sfig:AR}
(a) Degree of circular polarization $P_{\rm CP}$ for different aspect ratio of GNRs ranging from 1 to 5.
(b–f) spatial distributions inside the GNRs with aspect ratios 1, 2, 3, 4 and 5, respectively.}
\end{figure}

We also investigated the $P_{\rm CP}$ of the GNR dipole moment at 600 nm as a function of the GNR aspect ratio (AR), as shown in Fig.~\ref{sfig:AR}.
The case of $\mathrm{AR}=1$ corresponds to a sphere, resulting in a $P_{\rm CP}$ of zero. The $P_{\rm CP}$ reaches approximately 23\% at $\mathrm{AR}=2$.
Increasing the AR beyond this value does not lead to a further enhancement of DCP, ostensibly due to the fact that the mode settles into a wire mode which has no net DCP.


\section{Maximizing the DCP of emitted light}
A natural question regarding the results we report here is how large a degree of circular polarization can be achieved? A systematic exploration of this question over all experimental parameters is beyond the scope of the present paper. Instead we provide evidence below that $|P_{/rm CP}|$ can become close to unity for experimentally achievable parameters.

\begin{figure}
\centering
\includegraphics[width=0.9\linewidth]{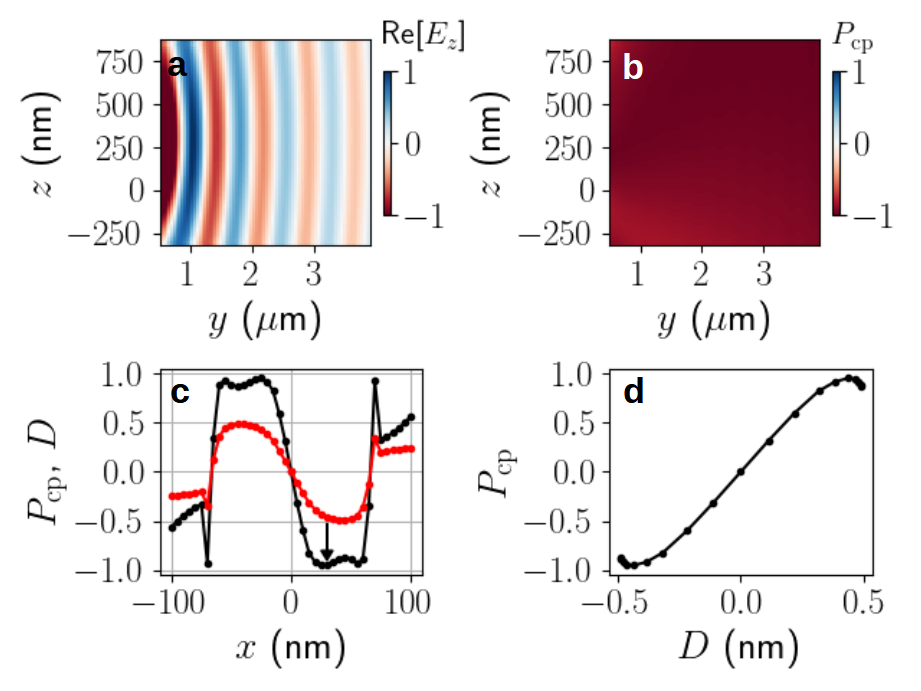}
\caption{\label{sfig:max} (a) Emission of light with almost perfect circular polarization. (b) DCP of emission shown in (a). (c) DCP of emitted light (black dots) and directionality (red dots) as a function of the dipole position. The arrow indicates the dipole position which gives the results in (a) and (b). (d) Degree of polarization $P_{\rm CP}$ as a function of directionality $D$. }
\end{figure}

In particular, as shown in Fig.~\ref{sfig:max}, using the same GNR dimensions as in the main paper, but lengthening the emission wavelength from 600 nm to 750 nm is enough to produce a DCP for the emitted light of $>0.9$ (Fig.~\ref{sfig:max}(a),(b)). Interestingly, the directionality achieved for this near maximal DCP only approaches 50$\%$ as seen in Figs.~\ref{sfig:max}(c),(d). The exact reason why a near perfect circular polarization does not correspond to a near perfect directionality is not yet completely clear, but is likely due to the modification of the fiber's evanescent field polarization due to the presence of the GNR.  

\section{Experimental methods}
Here we give more details of the experimental setup, methods and analysis as introduced in the main text
\subsection{Experimental setup}

Our experiments take place inside a scanning electron microscope (SEM, Carl-Zeiss SUPRA 40) with a homemade fiber feedthrough~\cite{abraham1998teflon}. An optical nanofiber fiber manufactured by a standard heat-and-pull method~\cite{birks2002shape} is mounted inside the SEM and spliced to the fiber feedthrough.

By further splicing the output fiber on the air-side to a 600 nm single mode fiber (SMF) and using 600 nm long pass filters (LPF) before detection of the signal at the SPCMs, as shown in Fig.~3(d) of the main text, we ensured that the detected CL was that coupled to the fundamental mode of the fiber. For this mode, the evanescent field extends approximately one wavelength into the surrounding vacuum, and has an elliptical polarization in the $x-z$ plane dependent on the propagation direction, as shown by the thick black arrows in Fig.~3(a) of the main text. This spin-momentum locking in the evanescent field allows the conversion of polarization of CL emitted by the GNR to directionality of light in the fiber, as LCP (RCP) emission couples more strongly to the $+x$ ($-x$) propagating mode.

\subsubsection{Details of the electron beam scan}
Fig.~3(a) of the main text depicts the GNR on the fiber surface, along with the path of the electron beam, and the polarization state of the evanescent field of the fiber mode for $\pm x$ propagation. Note that the electron beam penetrates a distance $\delta$ into the fiber which is taken to be $10$ nm for the GNR, and $20$ nm for the silica fiber~\cite{uemura2021probing}.

The electron beam scan was performed at a rate of approximately 30 lines per minute, typically at a magnification between 25 and 50 times depending on the purpose of the scan. The diameter of the electron beam is approximately 5 nm, although the effective resolution of the scan is principally decided by the scan rate and SPCM output sample rate.

\subsubsection{CL data analysis}
Raw CL data was recorded as a time sequence of photon counts by SPCM1 and SPCM2. This was then reconstructed to give a CL image by aligning the data from each line scan. The characteristic rise due to the fiber edge was used to align the data when necessary.  

We calculate the directionality by calculating the ratio 
\[D=\frac{I_1-I_2}{I_1 + I_2}\]
for each point in the CL image, where $I_1$ and $I_2$ are the SPCM1 and SPCM2 photon counts respectively.

We note that which SPCM should be considered 1 and which should be considered 2 may not be clear, or can easily be mixed up by accidentally swapping the input fibers. Luckily, it is always possible to decide which is 1 and which is 2 post-measurement, since the SPCM1 end of the GNR gives larger intensities on average at SPCM1 rather than SPCM2 and vice versa. This allows the correct sign of $D$ to be determined at analysis time for comparison with simulations.

An additional complication in calculating $D$ is that although the relative changes in intensity at each SPCM are determined by the directionality of coupling, the \emph{average} value of the intensity at each SPCM, which ideally should be the same, can depend on the details of individual fiber tapers, including the shape of the taper itself on each side of the nanofiber region, and any impurities introduced asymmetrically on the fiber during the GNR deposition process. It is thus necessary in general to scale the signals $I_1$ and $I_2$ so that their average values are the same.
In addition, for comparison with simulations, it is necessary to set the experimental directionality for excitement at the GNR center to its ideal value of zero. For some data sets, this condition is already true to a good approximation, but for data sets where the mean values of $I_1$ and $I_2$ differ, an offset can exist at the center position. We subtract any such offset before comparison of the 1D data (i.e. on a line through the GNR center) with simulations, so that the qualitative shape of the directionality can be compared between experiments and simulations. 


\bibliography{PlasmonCL}

\providecommand{\latin}[1]{#1}
\makeatletter
\providecommand{\doi}
  {\begingroup\let\do\@makeother\dospecials
  \catcode`\{=1 \catcode`\}=2 \doi@aux}
\providecommand{\doi@aux}[1]{\endgroup\texttt{#1}}
\makeatother
\providecommand*\mcitethebibliography{\thebibliography}
\csname @ifundefined\endcsname{endmcitethebibliography}
  {\let\endmcitethebibliography\endthebibliography}{}
\begin{mcitethebibliography}{38}
\providecommand*\natexlab[1]{#1}
\providecommand*\mciteSetBstSublistMode[1]{}
\providecommand*\mciteSetBstMaxWidthForm[2]{}
\providecommand*\mciteBstWouldAddEndPuncttrue
  {\def\EndOfBibitem{\unskip.}}
\providecommand*\mciteBstWouldAddEndPunctfalse
  {\let\EndOfBibitem\relax}
\providecommand*\mciteSetBstMidEndSepPunct[3]{}
\providecommand*\mciteSetBstSublistLabelBeginEnd[3]{}
\providecommand*\EndOfBibitem{}
\mciteSetBstSublistMode{f}
\mciteSetBstMaxWidthForm{subitem}{(\alph{mcitesubitemcount})}
\mciteSetBstSublistLabelBeginEnd
  {\mcitemaxwidthsubitemform\space}
  {\relax}
  {\relax}

\bibitem[Purcell(1946)]{Purcell}
Purcell,~E.~M. Proceedings of the American Physical Society. \emph{Phys. Rev.}
  \textbf{1946}, \emph{69}, 681(B10)\relax
\mciteBstWouldAddEndPuncttrue
\mciteSetBstMidEndSepPunct{\mcitedefaultmidpunct}
{\mcitedefaultendpunct}{\mcitedefaultseppunct}\relax
\EndOfBibitem
\bibitem[Aharonovich \latin{et~al.}(2016)Aharonovich, Englund, and
  Toth]{aharonovich2016solid}
Aharonovich,~I.; Englund,~D.; Toth,~M. Solid-state single-photon emitters.
  \emph{Nature photonics} \textbf{2016}, \emph{10}, 631--641\relax
\mciteBstWouldAddEndPuncttrue
\mciteSetBstMidEndSepPunct{\mcitedefaultmidpunct}
{\mcitedefaultendpunct}{\mcitedefaultseppunct}\relax
\EndOfBibitem
\bibitem[Unitt \latin{et~al.}(2005)Unitt, Bennett, Atkinson, Ritchie, and
  Shields]{unitt2005polarization}
Unitt,~D.; Bennett,~A.; Atkinson,~P.; Ritchie,~D.; Shields,~A. Polarization
  control of quantum dot single-photon sources via a dipole-dependent Purcell
  effect. \emph{Physical Review B—Condensed Matter and Materials Physics}
  \textbf{2005}, \emph{72}, 033318\relax
\mciteBstWouldAddEndPuncttrue
\mciteSetBstMidEndSepPunct{\mcitedefaultmidpunct}
{\mcitedefaultendpunct}{\mcitedefaultseppunct}\relax
\EndOfBibitem
\bibitem[Munsch \latin{et~al.}(2012)Munsch, Claudon, Bleuse, Malik, Dupuy,
  G{\'e}rard, Chen, Gregersen, and M{\o}rk]{munsch2012linearly}
Munsch,~M.; Claudon,~J.; Bleuse,~J.; Malik,~N.~S.; Dupuy,~E.;
  G{\'e}rard,~J.-M.; Chen,~Y.; Gregersen,~N.; M{\o}rk,~J. Linearly polarized,
  single-mode spontaneous emission in a photonic nanowire. \emph{Physical
  Review Letters} \textbf{2012}, \emph{108}, 077405\relax
\mciteBstWouldAddEndPuncttrue
\mciteSetBstMidEndSepPunct{\mcitedefaultmidpunct}
{\mcitedefaultendpunct}{\mcitedefaultseppunct}\relax
\EndOfBibitem
\bibitem[Zhu \latin{et~al.}(2014)Zhu, Zheng, Lin, Liu, and
  Jin]{zhu2014polarization}
Zhu,~Q.; Zheng,~S.; Lin,~S.; Liu,~T.-R.; Jin,~C. Polarization-dependent
  enhanced photoluminescence and polarization-independent emission rate of
  quantum dots on gold elliptical nanodisc arrays. \emph{Nanoscale}
  \textbf{2014}, \emph{6}, 7237--7242\relax
\mciteBstWouldAddEndPuncttrue
\mciteSetBstMidEndSepPunct{\mcitedefaultmidpunct}
{\mcitedefaultendpunct}{\mcitedefaultseppunct}\relax
\EndOfBibitem
\bibitem[Pfeiffer \latin{et~al.}(2018)Pfeiffer, Atkinson, Rastelli, Schmidt,
  Giessen, Lippitz, and Lindfors]{pfeiffer2018coupling}
Pfeiffer,~M.; Atkinson,~P.; Rastelli,~A.; Schmidt,~O.~G.; Giessen,~H.;
  Lippitz,~M.; Lindfors,~K. Coupling a single solid-state quantum emitter to an
  array of resonant plasmonic antennas. \emph{Scientific reports}
  \textbf{2018}, \emph{8}, 3415\relax
\mciteBstWouldAddEndPuncttrue
\mciteSetBstMidEndSepPunct{\mcitedefaultmidpunct}
{\mcitedefaultendpunct}{\mcitedefaultseppunct}\relax
\EndOfBibitem
\bibitem[Chandra \latin{et~al.}(2020)Chandra, Ahmed, and
  McCormack]{chandra2020polarization}
Chandra,~S.; Ahmed,~H.; McCormack,~S. Polarization-sensitive anisotropic
  plasmonic properties of quantum dots and Au nanorod composites. \emph{Optics
  Express} \textbf{2020}, \emph{28}, 20191--20204\relax
\mciteBstWouldAddEndPuncttrue
\mciteSetBstMidEndSepPunct{\mcitedefaultmidpunct}
{\mcitedefaultendpunct}{\mcitedefaultseppunct}\relax
\EndOfBibitem
\bibitem[Zhang \latin{et~al.}(2019)Zhang, Li, Wang, Tian, Chen, Fountaine,
  DiMarzio, Liu, Cotlet, and Gang]{zhang2019polarized}
Zhang,~H.; Li,~M.; Wang,~K.; Tian,~Y.; Chen,~J.-S.; Fountaine,~K.~T.;
  DiMarzio,~D.; Liu,~M.; Cotlet,~M.; Gang,~O. Polarized single-particle quantum
  dot emitters through programmable cluster assembly. \emph{ACS nano}
  \textbf{2019}, \emph{14}, 1369--1378\relax
\mciteBstWouldAddEndPuncttrue
\mciteSetBstMidEndSepPunct{\mcitedefaultmidpunct}
{\mcitedefaultendpunct}{\mcitedefaultseppunct}\relax
\EndOfBibitem
\bibitem[Sugawara \latin{et~al.}(2022)Sugawara, Xuan, Mitsumori, Edamatsu, and
  Sadgrove]{sugawara2022plasmon}
Sugawara,~M.; Xuan,~Y.; Mitsumori,~Y.; Edamatsu,~K.; Sadgrove,~M.
  Plasmon-enhanced single photon source directly coupled to an optical fiber.
  \emph{Physical Review Research} \textbf{2022}, \emph{4}, 043146\relax
\mciteBstWouldAddEndPuncttrue
\mciteSetBstMidEndSepPunct{\mcitedefaultmidpunct}
{\mcitedefaultendpunct}{\mcitedefaultseppunct}\relax
\EndOfBibitem
\bibitem[Shafi \latin{et~al.}(2023)Shafi, Yalla, and Nayak]{shafi2023bright}
Shafi,~K.~M.; Yalla,~R.; Nayak,~K.~P. Bright and polarized fiber in-line
  single-photon source based on plasmon-enhanced emission into nanofiber guided
  modes. \emph{Physical Review Applied} \textbf{2023}, \emph{19}, 034008\relax
\mciteBstWouldAddEndPuncttrue
\mciteSetBstMidEndSepPunct{\mcitedefaultmidpunct}
{\mcitedefaultendpunct}{\mcitedefaultseppunct}\relax
\EndOfBibitem
\bibitem[Chen \latin{et~al.}(2025)Chen, Wang, Si, Zhang, Yin, Chen, Lv, Tang,
  Zheng, Kivshar, \latin{et~al.} others]{chen2025observation}
Chen,~Y.; Wang,~M.; Si,~J.; Zhang,~Z.; Yin,~X.; Chen,~J.; Lv,~N.; Tang,~C.;
  Zheng,~W.; Kivshar,~Y., \latin{et~al.}  Observation of chiral emission
  enabled by collective guided resonances. \emph{Nature Nanotechnology}
  \textbf{2025}, 1--8\relax
\mciteBstWouldAddEndPuncttrue
\mciteSetBstMidEndSepPunct{\mcitedefaultmidpunct}
{\mcitedefaultendpunct}{\mcitedefaultseppunct}\relax
\EndOfBibitem
\bibitem[Ahn \latin{et~al.}(2024)Ahn, Le, Narushima, Yamanishi, Kim, Nam, and
  Okamoto]{ahn2024highly}
Ahn,~H.-Y.; Le,~K.~Q.; Narushima,~T.; Yamanishi,~J.; Kim,~R.~M.; Nam,~K.~T.;
  Okamoto,~H. Highly chiral light emission using plasmonic helicoid
  nanoparticles. \emph{Advanced Optical Materials} \textbf{2024}, \emph{12},
  2400699\relax
\mciteBstWouldAddEndPuncttrue
\mciteSetBstMidEndSepPunct{\mcitedefaultmidpunct}
{\mcitedefaultendpunct}{\mcitedefaultseppunct}\relax
\EndOfBibitem
\bibitem[Xie \latin{et~al.}(2025)Xie, Krasavin, Roth, and
  Zayats]{xie2025unidirectional}
Xie,~Y.; Krasavin,~A.~V.; Roth,~D.~J.; Zayats,~A.~V. Unidirectional chiral
  scattering from single enantiomeric plasmonic nanoparticles. \emph{Nature
  Communications} \textbf{2025}, \emph{16}, 1125\relax
\mciteBstWouldAddEndPuncttrue
\mciteSetBstMidEndSepPunct{\mcitedefaultmidpunct}
{\mcitedefaultendpunct}{\mcitedefaultseppunct}\relax
\EndOfBibitem
\bibitem[Lodahl \latin{et~al.}(2017)Lodahl, Mahmoodian, Stobbe, Rauschenbeutel,
  Schneeweiss, Volz, Pichler, and Zoller]{lodahl2017chiral}
Lodahl,~P.; Mahmoodian,~S.; Stobbe,~S.; Rauschenbeutel,~A.; Schneeweiss,~P.;
  Volz,~J.; Pichler,~H.; Zoller,~P. Chiral quantum optics. \emph{Nature}
  \textbf{2017}, \emph{541}, 473--480\relax
\mciteBstWouldAddEndPuncttrue
\mciteSetBstMidEndSepPunct{\mcitedefaultmidpunct}
{\mcitedefaultendpunct}{\mcitedefaultseppunct}\relax
\EndOfBibitem
\bibitem[Petersen \latin{et~al.}(2014)Petersen, Volz, and
  Rauschenbeutel]{petersen2014chiral}
Petersen,~J.; Volz,~J.; Rauschenbeutel,~A. Chiral nanophotonic waveguide
  interface based on spin-orbit interaction of light. \emph{Science}
  \textbf{2014}, \emph{346}, 67--71\relax
\mciteBstWouldAddEndPuncttrue
\mciteSetBstMidEndSepPunct{\mcitedefaultmidpunct}
{\mcitedefaultendpunct}{\mcitedefaultseppunct}\relax
\EndOfBibitem
\bibitem[Le~Feber \latin{et~al.}(2015)Le~Feber, Rotenberg, and
  Kuipers]{le2015nanophotonic}
Le~Feber,~B.; Rotenberg,~N.; Kuipers,~L. Nanophotonic control of circular
  dipole emission. \emph{Nature communications} \textbf{2015}, \emph{6},
  6695\relax
\mciteBstWouldAddEndPuncttrue
\mciteSetBstMidEndSepPunct{\mcitedefaultmidpunct}
{\mcitedefaultendpunct}{\mcitedefaultseppunct}\relax
\EndOfBibitem
\bibitem[Sch{\"a}ferling \latin{et~al.}(2012)Sch{\"a}ferling, Yin, and
  Giessen]{schaferling2012formation}
Sch{\"a}ferling,~M.; Yin,~X.; Giessen,~H. Formation of chiral fields in a
  symmetric environment. \emph{Optics Express} \textbf{2012}, \emph{20},
  26326--26336\relax
\mciteBstWouldAddEndPuncttrue
\mciteSetBstMidEndSepPunct{\mcitedefaultmidpunct}
{\mcitedefaultendpunct}{\mcitedefaultseppunct}\relax
\EndOfBibitem
\bibitem[Hashiyada \latin{et~al.}(2018)Hashiyada, Narushima, and
  Okamoto]{hashiyada2018imaging}
Hashiyada,~S.; Narushima,~T.; Okamoto,~H. Imaging chirality of optical fields
  near achiral metal nanostructures excited with linearly polarized light.
  \emph{ACS Photonics} \textbf{2018}, \emph{5}, 1486--1492\relax
\mciteBstWouldAddEndPuncttrue
\mciteSetBstMidEndSepPunct{\mcitedefaultmidpunct}
{\mcitedefaultendpunct}{\mcitedefaultseppunct}\relax
\EndOfBibitem
\bibitem[Hashiyada \latin{et~al.}(2019)Hashiyada, Narushima, and
  Okamoto]{hashiyada2019active}
Hashiyada,~S.; Narushima,~T.; Okamoto,~H. Active control of chiral optical near
  fields on a single metal nanorod. \emph{ACS Photonics} \textbf{2019},
  \emph{6}, 677--683\relax
\mciteBstWouldAddEndPuncttrue
\mciteSetBstMidEndSepPunct{\mcitedefaultmidpunct}
{\mcitedefaultendpunct}{\mcitedefaultseppunct}\relax
\EndOfBibitem
\bibitem[Chang \latin{et~al.}(2007)Chang, S{\o}rensen, Hemmer, and
  Lukin]{chang2007strong}
Chang,~D.~E.; S{\o}rensen,~A.~S.; Hemmer,~P.; Lukin,~M. Strong coupling of
  single emitters to surface plasmons. \emph{Physical Review B—Condensed
  Matter and Materials Physics} \textbf{2007}, \emph{76}, 035420\relax
\mciteBstWouldAddEndPuncttrue
\mciteSetBstMidEndSepPunct{\mcitedefaultmidpunct}
{\mcitedefaultendpunct}{\mcitedefaultseppunct}\relax
\EndOfBibitem
\bibitem[Alpeggiani \latin{et~al.}(2018)Alpeggiani, Bliokh, Nori, and
  Kuipers]{alpeggiani2018electromagnetic}
Alpeggiani,~F.; Bliokh,~K.; Nori,~F.; Kuipers,~L. Electromagnetic helicity in
  complex media. \emph{Physical review letters} \textbf{2018}, \emph{120},
  243605\relax
\mciteBstWouldAddEndPuncttrue
\mciteSetBstMidEndSepPunct{\mcitedefaultmidpunct}
{\mcitedefaultendpunct}{\mcitedefaultseppunct}\relax
\EndOfBibitem
\bibitem[Yang \latin{et~al.}(2023)Yang, Mou, Zapata, Reynier, Gallas, and
  Mivelle]{yang2023inverse}
Yang,~X.; Mou,~Y.; Zapata,~R.; Reynier,~B.; Gallas,~B.; Mivelle,~M. An inverse
  Faraday effect generated by linearly polarized light through a plasmonic
  nano-antenna. \emph{Nanophotonics} \textbf{2023}, \emph{12}, 687--694\relax
\mciteBstWouldAddEndPuncttrue
\mciteSetBstMidEndSepPunct{\mcitedefaultmidpunct}
{\mcitedefaultendpunct}{\mcitedefaultseppunct}\relax
\EndOfBibitem
\bibitem[Griffiths(2023)]{griffiths2023introduction}
Griffiths,~D.~J. \emph{Introduction to electrodynamics}; Cambridge University
  Press, 2023\relax
\mciteBstWouldAddEndPuncttrue
\mciteSetBstMidEndSepPunct{\mcitedefaultmidpunct}
{\mcitedefaultendpunct}{\mcitedefaultseppunct}\relax
\EndOfBibitem
\bibitem[Takahara \latin{et~al.}(1997)Takahara, Yamagishi, Taki, Morimoto, and
  Kobayashi]{takahara1997guiding}
Takahara,~J.; Yamagishi,~S.; Taki,~H.; Morimoto,~A.; Kobayashi,~T. Guiding of a
  one-dimensional optical beam with nanometer diameter. \emph{Optics letters}
  \textbf{1997}, \emph{22}, 475--477\relax
\mciteBstWouldAddEndPuncttrue
\mciteSetBstMidEndSepPunct{\mcitedefaultmidpunct}
{\mcitedefaultendpunct}{\mcitedefaultseppunct}\relax
\EndOfBibitem
\bibitem[Garc{\'\i}a~de Abajo(2010)]{garcia2010optical}
Garc{\'\i}a~de Abajo,~F.~J. Optical excitations in electron microscopy.
  \emph{Reviews of modern physics} \textbf{2010}, \emph{82}, 209--275\relax
\mciteBstWouldAddEndPuncttrue
\mciteSetBstMidEndSepPunct{\mcitedefaultmidpunct}
{\mcitedefaultendpunct}{\mcitedefaultseppunct}\relax
\EndOfBibitem
\bibitem[Polman \latin{et~al.}(2019)Polman, Kociak, and Garc{\'\i}a~de
  Abajo]{polman2019electron}
Polman,~A.; Kociak,~M.; Garc{\'\i}a~de Abajo,~F.~J. Electron-beam spectroscopy
  for nanophotonics. \emph{Nature materials} \textbf{2019}, \emph{18},
  1158--1171\relax
\mciteBstWouldAddEndPuncttrue
\mciteSetBstMidEndSepPunct{\mcitedefaultmidpunct}
{\mcitedefaultendpunct}{\mcitedefaultseppunct}\relax
\EndOfBibitem
\bibitem[Garcia~de Abajo and Di~Giulio(2021)Garcia~de Abajo, and
  Di~Giulio]{garcia2021optical}
Garcia~de Abajo,~F.~J.; Di~Giulio,~V. Optical excitations with electron beams:
  challenges and opportunities. \emph{ACS photonics} \textbf{2021}, \emph{8},
  945--974\relax
\mciteBstWouldAddEndPuncttrue
\mciteSetBstMidEndSepPunct{\mcitedefaultmidpunct}
{\mcitedefaultendpunct}{\mcitedefaultseppunct}\relax
\EndOfBibitem
\bibitem[Novotny and Hecht(2012)Novotny, and Hecht]{novotny2012principles}
Novotny,~L.; Hecht,~B. \emph{Principles of nano-optics}; Cambridge university
  press, 2012\relax
\mciteBstWouldAddEndPuncttrue
\mciteSetBstMidEndSepPunct{\mcitedefaultmidpunct}
{\mcitedefaultendpunct}{\mcitedefaultseppunct}\relax
\EndOfBibitem
\bibitem[Le~Kien \latin{et~al.}(2004)Le~Kien, Liang, Hakuta, and
  Balykin]{le2004field}
Le~Kien,~F.; Liang,~J.; Hakuta,~K.; Balykin,~V. Field intensity distributions
  and polarization orientations in a vacuum-clad subwavelength-diameter optical
  fiber. \emph{Optics Communications} \textbf{2004}, \emph{242}, 445--455\relax
\mciteBstWouldAddEndPuncttrue
\mciteSetBstMidEndSepPunct{\mcitedefaultmidpunct}
{\mcitedefaultendpunct}{\mcitedefaultseppunct}\relax
\EndOfBibitem
\bibitem[Sugawara \latin{et~al.}(2020)Sugawara, Mitsumori, Edamatsu, and
  Sadgrove]{sugawara2020optical}
Sugawara,~M.; Mitsumori,~Y.; Edamatsu,~K.; Sadgrove,~M. Optical detection of
  nano-particle characteristics using coupling to a nano-waveguide.
  \emph{Optics Express} \textbf{2020}, \emph{28}, 18938--18945\relax
\mciteBstWouldAddEndPuncttrue
\mciteSetBstMidEndSepPunct{\mcitedefaultmidpunct}
{\mcitedefaultendpunct}{\mcitedefaultseppunct}\relax
\EndOfBibitem
\bibitem[Uemura \latin{et~al.}(2021)Uemura, Irita, Homma, and
  Sadgrove]{uemura2021probing}
Uemura,~Y.; Irita,~M.; Homma,~Y.; Sadgrove,~M. Probing the local density of
  states near the diffraction limit using nanowaveguide-collected cathode
  luminescence. \emph{Physical Review A} \textbf{2021}, \emph{104},
  L031504\relax
\mciteBstWouldAddEndPuncttrue
\mciteSetBstMidEndSepPunct{\mcitedefaultmidpunct}
{\mcitedefaultendpunct}{\mcitedefaultseppunct}\relax
\EndOfBibitem
\bibitem[Coenen \latin{et~al.}(2014)Coenen, Bernal~Arango, Femius~Koenderink,
  and Polman]{coenen2014directional}
Coenen,~T.; Bernal~Arango,~F.; Femius~Koenderink,~A.; Polman,~A. Directional
  emission from a single plasmonic scatterer. \emph{Nature communications}
  \textbf{2014}, \emph{5}, 3250\relax
\mciteBstWouldAddEndPuncttrue
\mciteSetBstMidEndSepPunct{\mcitedefaultmidpunct}
{\mcitedefaultendpunct}{\mcitedefaultseppunct}\relax
\EndOfBibitem
\bibitem[Lum()]{Lumerical}
Ansys Lumerical FDTD\relax
\mciteBstWouldAddEndPuncttrue
\mciteSetBstMidEndSepPunct{\mcitedefaultmidpunct}
{\mcitedefaultendpunct}{\mcitedefaultseppunct}\relax
\EndOfBibitem
\bibitem[Haynes(2016)]{haynes2016crc}
Haynes,~W.~M. \emph{CRC handbook of chemistry and physics}; CRC press,
  2016\relax
\mciteBstWouldAddEndPuncttrue
\mciteSetBstMidEndSepPunct{\mcitedefaultmidpunct}
{\mcitedefaultendpunct}{\mcitedefaultseppunct}\relax
\EndOfBibitem
\bibitem[Zarraoa \latin{et~al.}(2019)Zarraoa, Gonz{\'a}lez, and
  Paulo]{zarraoa2019imaging}
Zarraoa,~L.; Gonz{\'a}lez,~M.~U.; Paulo,~{\'A}.~S. Imaging low-dimensional
  nanostructures by very low voltage scanning electron microscopy:
  ultra-shallow topography and depth-tunable material contrast.
  \emph{Scientific Reports} \textbf{2019}, \emph{9}, 16263\relax
\mciteBstWouldAddEndPuncttrue
\mciteSetBstMidEndSepPunct{\mcitedefaultmidpunct}
{\mcitedefaultendpunct}{\mcitedefaultseppunct}\relax
\EndOfBibitem
\bibitem[Abraham and Cornell(1998)Abraham, and Cornell]{abraham1998teflon}
Abraham,~E.~R.; Cornell,~E.~A. Teflon feedthrough for coupling optical fibers
  into ultrahigh vacuum systems. \emph{Applied optics} \textbf{1998},
  \emph{37}, 1762--1763\relax
\mciteBstWouldAddEndPuncttrue
\mciteSetBstMidEndSepPunct{\mcitedefaultmidpunct}
{\mcitedefaultendpunct}{\mcitedefaultseppunct}\relax
\EndOfBibitem
\bibitem[Birks and Li(2002)Birks, and Li]{birks2002shape}
Birks,~T.~A.; Li,~Y.~W. The shape of fiber tapers. \emph{Journal of lightwave
  technology} \textbf{2002}, \emph{10}, 432--438\relax
\mciteBstWouldAddEndPuncttrue
\mciteSetBstMidEndSepPunct{\mcitedefaultmidpunct}
{\mcitedefaultendpunct}{\mcitedefaultseppunct}\relax
\EndOfBibitem
\end{mcitethebibliography}

\end{document}